\documentclass[twocolumn,secnumarabic,amssymb, amsmath,nobibnotes, aps, pre]{revtex4}
\usepackage{psfrag,pst-grad,amsmath,amssymb,pstricks,color}
\usepackage{graphicx}
\usepackage{mathrsfs}
\usepackage{bm}
\usepackage{natbib}
\begin{document}
\title{Global first passage times on fractal lattices}
\author{C.P. Haynes, A.P. Roberts}
\affiliation{School of Physical Sciences, University of Queensland, Queensland 4072, Australia}

\begin{abstract}
The global first passage time density of a network is the probability that a random walker released at a random site arrives at an absorbing trap at time $T$. We find simple expressions for the mean global first passage time $\langle T \rangle$ for five fractals: the $d$-dimensional Sierpinski gasket, T-fractal, hierarchical percolation model, Mandelbrot-Given curve and a deterministic tree. We also find an exact expression for the second moment $\langle T^2 \rangle$ and show that the variance of the first passage time, $\textrm{Var}(T)$, scales with the number of nodes within the fractal $N$ such that $\textrm{Var}(T) \sim N^{4/\bar{d}}$, where $\bar{d}$ is the spectral dimension. 
\end{abstract}

\maketitle

\section{Introduction}
The study of diffusion on statistically self-similar disordered media (e.g. percolation clusters and diffusion limited aggregates) is simplified by modeling their structure by deterministic fractals \cite{Yuven}. Although scaling theory has been central in studying diffusive processes on a variety of deterministic fractals \cite{Havlin}, it has been noted that scaling laws do not provide a complete picture of dynamic phenomena on fractals \cite{Giona}, and exact relationships are useful. 

Lately there has been interest in calculating the mean time $\langle T\rangle$ taken for a walker released from a randomly chosen node on a fractal to arrive at a trap. We call the underlying probability density of $T$ the global first passage time (GFPT). It is found by a simple average of the node-to-trap first passage times over the entire fractal. 
The mean of the GFPT was first obtained for the Sierpinski gasket lattice by Kozak and Balakrishnan \cite{Kozak2,Kozak}, and an analogous study has been recently performed for the T-fractal\cite{Agliara}. The mean $\langle T \rangle$ has been associated \cite{Agliara} with important questions such as the exploration of a random walker on a fractal lattice \cite{Condamin}, providing further motivation for the study of global first passage times.

In general, deterministic fractals have played a key role in explaining diffusion in disordered materials because their properties can be exactly studied. In this paper we use a renormalization method to derive an exact expression for the full moment generating function of the GFPT. The method is exact, and can be applied to any finitely ramified deterministic fractal, making the exploration of higher moments of the GFPT possible. There is interest in higher moments of random walk quantities \cite{Kopelman,Hughes} as such results give further insight into the nature of transport on fractals. Knowledge of these moments is also useful to investigate the variance of the bi-molecular reaction $A+B \rightarrow A$ \cite{Rass}.

The paper is organized as follows:
In \S 2 we formulate the problem of continuum diffusion on networks of pipes as an algebraic system of equations involving Laplace transforms of the fluxes and concentrations at each node. 
We show how the equations are solved on a $d$-dimensional Sierpinski gasket lattice using a renormalization procedure. In \S 3, we derive the general formula for the Laplace transform of the GFPT density for a trap at an arbitrary node on the gasket. We demonstrate our method for the apex of the $n$th iteration of the $d$-dimensional gasket. Our results, for the mean of the GFPT, agree with prior work \cite{Kozak2,Kozak}, and are extended to show that, for any given trap, there is an exact analytic expression for the moment generating function of the GFPT.

In \S 4, we demonstrate that the method can be used to calculate the GFPT moment generating function for at least one node on any structure provided that a renormalization procedure exists. A general formula is provided for the GFPT moment generating function for a finitely ramified deterministic fractal with two end nodes. As examples we consider the Mandelbrot-Given curve, deterministic tree, hierarchical percolation model and the T-tree \cite{Agliara}. The calculation of higher moments is demonstrated for the T-tree in \S 5. 
 
\section{continuum diffusion on networks}\label{sec:section1} 
We begin by solving the diffusion equation  
\begin{equation} \label{eq:chap1}
D  {\frac {\partial ^{2}}{\partial {x}^{2}}}C \left( x,
t \right) ={\frac {\partial }{\partial t}}C \left( x,t \right)
\end{equation}
on a bar of length $L$ with the non-homogeneous boundary conditions 
\begin{displaymath}
C \left( 0,t \right) = P_1 \left( t \right),  
\quad C \left( L,t \right)  =P_2\left( t \right)  
\end{displaymath}
and initial condition $C \left( x,0 \right) = 0$.
Taking Laplace transforms of \eqref{eq:chap1} and solving using the boundary and initial conditions we find  
\begin{equation} \label{eq:chap3}
c \left( x,s \right)  =  p_2(s)\,\frac{\sinh \left( x\sqrt {{\frac {s}{D}}}
 \right) }{\sinh \left(L\sqrt {{\frac {s}{D}}}\right)}+ p_1(s)\, \frac{\sinh \left(  \left( L-x \right) \sqrt {{
\frac {s}{D}}} \right) }{ \sinh \left( L\sqrt {{\frac {s}{D}}}\right)},
\end{equation}
where we denote Laplace transformed functions by lower case letters;
\begin{displaymath}
c(x,s) = \mathscr{L}(C(x,t)) \equiv \int_0^\infty C\left(x,t \right) e^{-st}dt.
\end{displaymath}

The flux entering the bar at the point $x = 0$ is  
$F_1 = -D\frac{\partial }{\partial x} C \left(x,t \right)\Big{|}_{x=0}$ and the flux entering the bar at $x = L$ is $F_2 = D\frac{\partial }{\partial x} C \left(x,t \right)\Big{|}_{x=L}$. For networks it is useful to express these fluxes in terms of the concentrations $p_i$ at either end. This gives rise to the equations
\begin{equation} \label{eq:chap4}
g\left[ \begin {array}{c} {f_1 }\\\noalign{\medskip}{ f_2}
\end {array} \right] =\left[ \begin {array}{cc} 1 & - { \textrm{sech}} \left( \sqrt {{\frac{s}{D}}}L \right) 
\\\noalign{\medskip}-{ \textrm{sech}} \left( \sqrt {{
\frac {s}{D}}}L \right) & 1 \end {array} \right]  \left[ \begin {array}
{c} { p_1}\\\noalign{\medskip}{ p_2}\end {array} \right],
\end{equation}
where $g =  \tanh \left( \sqrt {{\frac{s}{D}}}L \right)/\sqrt{sD}$. We call the above a flux-concentration matrix. For  simplicity we set $D$ and $L$ to be unity.

Using these relations, diffusion on an arbitrary network can be formulated as a system of algebraic equations. To link two or more bars at a node we set the concentrations to be equal. As an example the system for the 1st iteration of the Sierpinski gasket lattice shown in Figure 1 (a) is
\begin{equation}\label{eq:chap12}
g\left[ \begin {array}{c} { f_1}\\\noalign{\medskip}{f_2}
\\\noalign{\medskip}{ f_3}\\\noalign{\medskip}{ f_4}
\\\noalign{\medskip}{f_5}\\\noalign{\medskip}{ f_6}
\end {array} \right] 
= 
\left[ \begin {array}{cccccc} 2\,a_0&-b_0&-b_0&0&0&0\\\noalign{\medskip}-b_0&4
\,a_0&-b_0&-b_0&-b_0&0\\\noalign{\medskip}-b_0&-b_0&4\,a_0&0&-b_0&-b_0\\\noalign{\medskip}0&-b_0&0
&2\,a_0&-b_0&0\\\noalign{\medskip}0&-b_0&-b_0&-b_0&4\,a_0&-b_0\\\noalign{\medskip}0&0&-b_0
&0&-b_0&2\,a_0\end {array} \right] 
\left[ \begin {array}{c} { p_1}\\\noalign{\medskip}{ p_2}
\\\noalign{\medskip}{ p_3}\\\noalign{\medskip}{ p_4}
\\\noalign{\medskip}{ p_5}\\\noalign{\medskip}{ p_6}
\end {array} \right]. 
\end{equation}
Here $a_0 = 1$ and $b_0 = \textrm{sech}\left(\sqrt {s}\right)$. 

\subsection{Initial and boundary conditions}\label{icbc}
The $f_i$ in (\ref{eq:chap12}) are understood to represent the total flux entering node $i$. For mass conservation $f_i = 0$, but if we take an instantaneous unit source at node $i$, then the flux entering the node is $F_i = \delta(t)$, where $\delta(t)$ is the Dirac $\delta$ function. This means that $f_i = \mathscr{L}(F_i) = \mathscr{L}\left( \delta(t) \right) = 1$. The matrix has a similar form to a simple finite difference scheme for modeling diffusion/probability on a lattice. Prior studies \cite{Koplik,Ark,Redner} have also formulated the problem of diffusion on networks in terms of the Laplace transform of the concentration and flux at a node. Here, linear algebra is used to extend these ideas. 

It is useful to illustrate the formulation of two different but closely related problems. If we take ${\bf f}^T=(0,0,0,0,0,1)$ the solution ${\bf p}^T= (p_1,p_2,p_3,p_4,p_5,p_6)$ provides the Laplace transforms of the 
concentrations at the nodes if an instantaneous source is released at node 6, and mass is conserved at every node (including node 6).
Next consider taking ${\bf f}^T=(1,0,0,0,0,f_6^\prime)$ and ${\bf p}^T=(p_1^\prime,p_2^\prime,p_3^\prime,p_4^\prime,p_5^\prime,0)$ and solving the 6 equations for $p_i^\prime$ ($i=1,\ldots,5$)  and $f_6^\prime$. In this case the equations correspond to an instantaneous source at node 1, a homogenous Dirichlet condition (i.e. a trap) at node 6, and mass conservation at nodes $i = 1,\ldots,5$.
The quantity $f_6^\prime$ is just the flux entering at the ``trap". It can be shown~\cite{thesis}  that
this flux can be calculated from $p_1$ and $p_6$ derived in the former problem, the result being $f_6^\prime= -p_1/p_6$. Note the negative sign arises from our convention for defining the flux.
The result (a convolution in time) can also be derived using a continuum random walker argument. We use a generalization of
this result in \S3.

\begin{figure}
\begin{center}
\includegraphics[width = 0.4\textwidth]{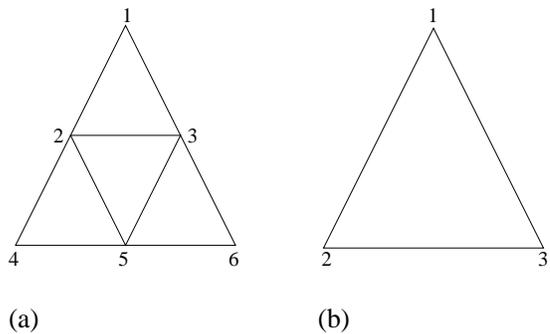}
\end{center}
\caption{(a) 1st iteration of the Sierpinski Gasket lattice. (b) 0th iteration of the Sierpinski gasket lattice. The labeling used in the above (a) and (b) implies that the left corner of the $n$th iteration of the $2$-$d$ Sierpinski gasket lattice
is denoted by $3\left(3^n+1\right)/2 - 2^n$, the right corner by $3\left(3^n+1\right)/2$ and the apex by 1.}
\label{fig:fig1}
\end{figure}

\subsection{First passage times and concentrations}\label{sec:section22}
In order to analyze the flux-concentration matrix for higher order generations of the Sierpinski gasket we apply a renormalization method. We first show how the $6 \times 6$ system \eqref{eq:chap12} can be reduced to a $3 \times 3$ system associated with the 0th order Sierpinski gasket (See Figure 1 (b))
\begin{equation}\label{eq:chap9}
g\left[ \begin {array}{c} { f_1}\\\noalign{\medskip}{ f_2}
\\\noalign{\medskip}{ f_3}\end {array} \right] 
=
\left[ \begin {array}{ccc} 2\,a_0&-b_0&-b_0\\\noalign{\medskip}-b_0& 2\,a_0&-b_0
\\\noalign
{\medskip}-b_0&-b_0&2\,a_0\end {array} \right] 
\left[ \begin {array}{c} { p_1}\\\noalign{\medskip}{ p_2}
\\\noalign{\medskip}{ p_3}\end {array} \right].
\end{equation}
To do this we consider the three equations associated with nodes 2,3 and 5 in equation \eqref{eq:chap12} 
\begin{displaymath}
\left[ \begin {array}{c} 0\\\noalign{\medskip} 0
\\\noalign{\medskip} 0
\end {array} \right] 
= 
\left[ \begin {array}{cccccc} -b_0&4
\,a_0&-b_0&-b_0&-b_0&0\\\noalign{\medskip}-b_0&-b_0&4\,a_0&0&-b_0&-b_0\\\noalign{\medskip}
0&-b_0&-b_0&-b_0&4\,a_0&-b_0\end {array} \right] 
\left[ \begin {array}{c} { p_1}\\\noalign{\medskip}{ p_2}
\\\noalign{\medskip}{ p_3}\\\noalign{\medskip}{ p_4}
\\\noalign{\medskip}{ p_5}\\\noalign{\medskip}{ p_6},
\end {array} \right] 
\end{displaymath}
where we have set $f_2=0,f_3=0$ and $f_5 = 0$. Solving this under determined system gives
\begin{equation} \label{eq:chap23}
\left[ \begin {array}{c} {p_2}\\\noalign{\medskip}{p_3}
\\\noalign{\medskip}{ p_5}\end {array} \right] 
= \frac{1}{(4\rho_0 + 1)(2\rho_0- 1)}
\left[ \begin {array}{ccc} \frac{2}{\rho_0} & \frac{2}{\rho_0} & 1 \\\noalign{\medskip} \frac{2}{\rho_0} & 1 & \frac{2}{\rho_0}
\\\noalign{\medskip} 1 & \frac{2}{\rho_0} & \frac{2}{\rho_0} \end {array} \right] 
\left[ \begin {array}{c} { p_1}\\\noalign{\medskip}{ p_4}
\\\noalign{\medskip}{ p_6}\end {array} \right],
\end{equation}
where $\rho_0 = \frac{a_0}{b_0}$. 
Eliminating $p_2,p_3$ and $p_5$ from \eqref{eq:chap12} by using \eqref{eq:chap23} gives
\begin{displaymath}
g\left[ \begin {array}{c} {f_1}\\\noalign{\medskip}{ f_4}
\\\noalign{\medskip}{f_6}\end {array} \right] 
=
\left[ \begin {array}{ccc} 2\,a_1&-b_1&-b_1\\\noalign{\medskip}-b_1& 2\,a_1&-b_1
\\\noalign
{\medskip}-b_1&-b_1&2\,a_1\end {array} \right] 
\left[ \begin {array}{c} { p_1}\\\noalign{\medskip}{ p_4}
\\\noalign{\medskip}{ p_6}\end {array} \right],
\end{displaymath}
with
\begin{displaymath}
b_{{1}}={\frac {a_{{1}}{b_{{0}}}^{2}}{a_{{0}} \left( 4\,a_{{0}}-3\,b_{
{0}} \right) }} = {\frac {{b_{{0}}}^{2} \left( 2\,a_{{0}}+b_{{0}} \right) }{ \left( 2\,a
_{{0}}-b_{{0}} \right)  \left( 4\,a_{{0}}+b_{{0}} \right) }}.
\end{displaymath}
Note that this flux-concentration matrix is of the same form as a flux-concentration matrix \eqref{eq:chap9} of the triangle in Figure 1 (b). 
Continuing, we find that the system of equations for the $n$th iteration of the Sierpinski gasket has the form \eqref{eq:chap9} with $a_0$ and $b_0$ replaced by
\begin{equation}\label{eq:chap26}
a_{n} = \rho_nb_n = {\rho_n \frac {b_{n-1} \left( 2\,\rho_{n-1}+1\right)  }{ \left( 4\,\rho_{n-1}+1 \right)  \left( 2\,\rho_{n-1}-1 \right) }}.
\end{equation}

Here $\rho_n$ is the renormalization of the inverse FPT equation on a 2-$d$ Sierpinski gasket, which satisfies 
\begin{displaymath}
\rho_{n} =  \rho_{n-1}\left(4\rho_{n-1}-3\right).
\end{displaymath}
A detailed discussion of the equation can be found in~\cite{Van2,Van,Grabner}.

The recurrence relationship for the $d$-dimensional Sierpinski gasket lattice is
\begin{equation}\label{eq:chap28}
a_{n} = \rho_nb_{n} = {\rho_n\frac { \left(  d\rho_{n-1}+  1\right) b_{n-1} }{ \left( d\,\rho_{n-1}- \left( d-1 \right)\right)  \left( 2d\, \rho_{n-1}+\left(3-d\right)\right) }},
\end{equation}
where 
\begin{equation}\label{eq:chap29}
\rho_n\left(d\right) =  2d\rho^2_{n-1} -3\left(d-1\right)\rho_{n-1} + \left(d-2\right)
\end{equation}
is the renormalization of the inverse FPT equation in $d$-dimensions. 
Equation \eqref{eq:chap29} has been derived in \cite{Yuste,Barlow} however the above recurrence relationships 
\eqref{eq:chap26} and \eqref{eq:chap28} are new, and allow the concentration to be calculated everywhere within the $n$th iteration of the $d$-dimensional Sierpinski gasket lattice given that an instantaneous unit source is released at a corner of the structure. 

Before proceeding, a comment needs to be made about equation \eqref{eq:chap23}. The relation between the concentrations calculated in \eqref{eq:chap23} holds for any iteration. By this it is understood that if $p_2,p_3,p_5$ can be expressed in terms of $p_1,p_4,p_6$ through \eqref{eq:chap23} using $\rho_0$ for the 1st iteration of the Sierpinski gasket, then $p_4,p_6,p_{13}$ can be expressed in terms of $p_1,p_{11},p_{15}$ through \eqref{eq:chap23} using $\rho_1$ on the 2nd iteration of the 2-$d$ Sierpinski gasket. In our notation (Figure 1),  node $13$ is the mid base node equidistant to the nodes $11$ and $15$, with the other nodes having a similar labeling to Figure 1. The fact that the concentrations at the nodes can be expressed through \eqref{eq:chap23} will be used when we derive a similar relation that holds for the concentrations at the nodes of other fractal structures in \S4.    

\section{First passage time}\label{sec:section3}
Let $F_{kj}(t)$ be the first passage time for a random walker released at node $j$ to reach node $k$. The GFPT is given by
$$
F_{k}(t) = \frac{1}{N}\sum_{j=1,j\neq k}^{N+1} F_{kj}(t)
$$
where $N+1$ is the total number of nodes and $N$ the number of starting nodes on the fractal. 
Let $p_{kj,n}$ be the concentration at a node $k$ given that an instantaneous unit source is released at a node $j$, and there is no loss of mass at all nodes (the first problem discussed in \S\ref{icbc}). The additional subscript $n$ denotes the iteration of the fractal. As in \S\ref{icbc} we consider a complementary problem where a trap is placed at node $k$ and a unit source is released at node $j$. The flux ``entering'' at $k$ due to the source at $j$ is 
\begin{displaymath}
f_{kj,n} = -\frac{p_{jk,n}}{p_{kk,n}}.
\end{displaymath}
 The Laplace transform of the GFPT for node $k$ is
\begin{equation}\label{eq:chap44}
f_{k,n} =\frac{1}{N}(-1 + \sum_{j=1}^{N+1} \big|f_{jk,n}\big|) = \frac{1}{N}\left(-1+ \frac{\sum_{j=1}^{N+1} p_{jk,n}}{p_{kk,n}}\right).
\end{equation}
The mean of the GFPT is $$\langle T_{k,n} \rangle = \int_0^\infty tF_{k}(t) dt = - \frac{df_{k,n}(s) }{ds}\Big|_{s=0}.$$ 
%
The form of $f_{k,n}$ \eqref{eq:chap44} is simplified for the Sierpinski gasket through the use of conservation of probability.
We integrate equation (\ref{eq:chap3}) for every pipe of the gasket and sum to find
\begin{equation}\label{eq:chap45}
\frac {d \left(\alpha_{k,n}+2\,{\beta_{k,n}} \right)  \left( \cosh
 \left( \sqrt{s} \right) -1 \right)}{\sqrt {s} 
\sinh \left( \sqrt {s} \right) }=\frac{1}{s}.
\end{equation} 
Here $\alpha_{k,n}$ is the sum of  concentrations at the corner nodes and the apex for the closed Sierpinski gasket and $\beta_{k,n}$ is the sum of the concentrations at the other $\left(d+1\right)\left(\left(d+1\right)^{n} - 1\right)/2$ nodes. Note that \eqref{eq:chap45} is true because there are $d$ pipes stemming from the corner nodes and there are $2d$ pipes stemming from each interior node. 
\begin{figure*}
\begin{center}
\includegraphics[width = 0.9\textwidth]{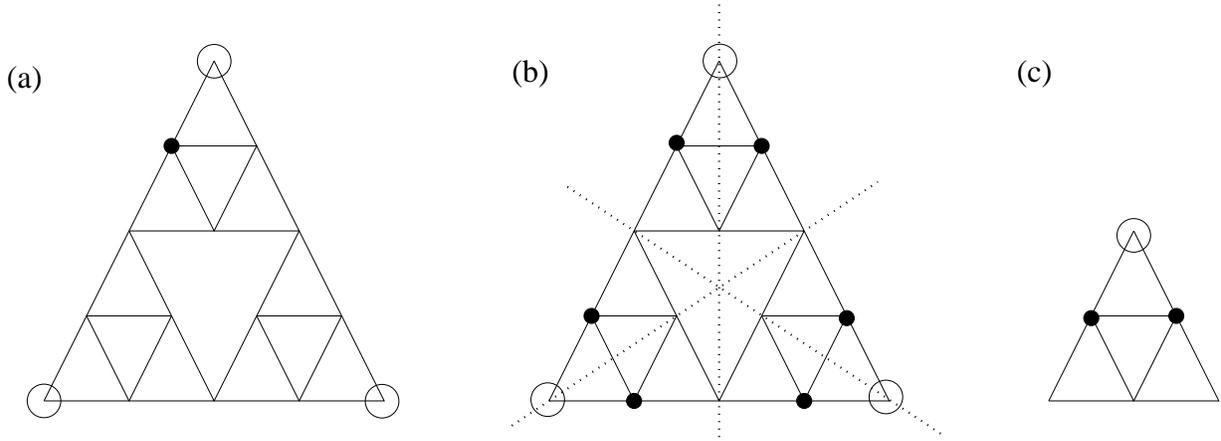}
\end{center}
\caption{Illustration of the method of images argument used to prove the result $\alpha_{k,n} = p_{1k,n-1}$. The filled circles denote sources and the open circles are the concentrations of interest.} 
\label{fig:fig7}
\end{figure*}

Using the method of images we show that $\alpha_{k,n} = p_{1k,n-1}$. Here we have identified the apex of the $n$th iteration of the $d$-dimensional Sierpinski gasket by the label 1. We consider the release of a walker at point $k = 2$ of the 2nd iteration of the 2-d gasket (see Figure \ref{fig:fig7} (a)) to show the result, but the extension to arbitrary $k$ and higher dimensions is clear.  
We first place an additional 5 sources at the image points of $k=2$ in the three lines of symmetry of the gasket (Figure \ref{fig:fig7} (b)). If we denote the sum of concentrations at the corner and apex nodes by $\alpha_{2,2}'$, then it can easily be seen (through symmetry) that $\alpha_{2,n} = \alpha_{2,2}'/6$. Now, by symmetry, there will be no flux between the three 1st iteration gaskets used to construct the 2nd iteration, and each problem (see Figure \ref{fig:fig7} (c)) can be treated alone. It is clear that the concentration at the apex of the reduced problem is $2p_{12,1}$. Therefore, $ \alpha_{2,2}' = 3(2p_{12,1})$ and finally $\alpha_{2,2} = p_{12,1}$. 
For a source at a point which lies on a line of symmetry (such as a corner),  only two copies of the source need to be considered. 
In the above argument $k$ is limited to values which appear in the $n-1$th iteration. All other points in the $n$th iteration have an image point corresponding to an acceptable $k$ value.

Using (\ref{eq:chap45}) reduces the GFPT (\ref{eq:chap44}) to
\begin{equation}\label{eq:chap46}
f_{k,n} = \frac{1}{2N}\left( {\frac {p_{kk,n}^{-1}\sinh \left( \sqrt {s} \right) 
}{d\sqrt{s} \left( \cosh \left( \sqrt {s} \right) -1 \right) 
}}+\frac{\alpha_{k,n}}{p_{kk,n}}-2 \right),
\end{equation}
where $N\left(d\right) = \left(\left(d+1\right)^{n+1}+d-1\right)/2$ is the number of starting nodes. 
Setting $\alpha_{k,n} = p_{1k,n-1}$ in \eqref{eq:chap46} gives
\begin{equation}\label{eq:chap47}
f_{k,n} = \frac{1}{2N}\left( {\frac {p_{kk,n}^{-1}\sinh \left( \sqrt {s} \right) 
}{d\sqrt{s} \left( \cosh \left( \sqrt {s} \right) -1 \right) 
}}+\frac{{ p_{1k,n-1}}}{p_{kk,n}}-2 \right).
\end{equation}

As an example, consider a trap at the apex which requires evaluation of $p_{11,n}$, which is provided in \eqref{eq:append4} of Appendix \ref{sec:append1}. After simplification the moment generating function of the GFPT for a trap at the apex is
\begin{equation}\label{eq:chap49}
f_{1,n} = \frac 1{2N} \left(\frac{d\rho_n +1}{d\rho_n-(d-1)} -2 + \prod _{j=0}^{n}\frac{d\rho_j +1}{d\rho_j -(d-1)} \right).
\end{equation}

\subsection{Discrete random walker}
Note that the diffusion equations and first passage times considered above correspond to a continuous-space/continuous-time random walker. We have chosen to formulate the problem in this way because it makes the working significantly simpler. For example, (\ref{eq:chap44}) applies to all networks, while it must be modified to account for local coordination numbers in the discrete formulation. From $f_{k,n}(s)$ we can find results for discrete-time/discrete-space walkers using the simple transformation discussed below.

Let $F^*_{kj}(m)$ denote the probability that a walker released at $j$ arrives at $k$ on the $m$th step for the
first time, so the GFPT is given by  $$F^*_{k}(m) = \frac{1}{N}\sum_{j=1,j\neq k}^{N+1} F^*_{jk}(m).$$
Here, and below, we denote quantities associated with a discrete walker by an asterisk.
The moment generating function of $F^*_{kj}(m)$ is defined as
$$
f^*_{kj}(\lambda) =\sum_{m=0}^\infty F^*_{kj}(m)e^{m\lambda}.
$$

$f^*_{kj}(\lambda)$ is the counter part of $f_{kj}(s)$ descibed above (the iteration for the fractal is irrelevant here). 
The two quantities are related by 
$$
f^*_{kj}(\lambda)= f_{kj}\left(\left(\textrm{arsech}(e^{\lambda})\right)^{2}\right).
$$
For example, on a single bar $T^*_{12}(m)= \delta_{m1}$  , so $f^*_{12}(\lambda)= e^{\lambda}$ and for a continuum walker $f_{12}(s)= 1/\rho_0(s) = \textrm{sech}\left(\sqrt{s}\right)$. A derivation of the transformation is given in Appendix \ref{sec:append2}. 

Thus, the moment generating function of the GFPT for a discrete walker is given by
\begin{equation}\label{eq:final1}
f^*_{k}(\lambda) = f_{k}\left(\left(\textrm{arsech}(e^{\lambda})\right)^{2}\right),
\end{equation}
so we can write
 $$\langle T^{*}_k \rangle = \sum_{m=0}^{\infty} mF^*_{k}(m) = \frac{df^*_k(\lambda) }{d\lambda}\Big|_{\lambda = 0}.$$ Since the location of the trap is generally clear, we simply write the mean as $\langle T^{*} \rangle$. To explicitly calculate $F^{*}_k(m)$, we form the probability generating function $f^*_{k}(\ln(z))$ and note that $F^*_k(m)$ is the coefficient of the $m$th term of the Taylor series expansion of $f^*_{k}(\ln(z))$ about $z=0$.

For the remainder of the paper, we first calculate $f_{k}(s)$, and then use \eqref{eq:final1} to determine  $f^*_{k}(\lambda)$. For the Sierpinski gasket example, a Taylor series exansions of the moment generating function $f^*_{k}(\lambda)$ (\eqref{eq:chap49} in \eqref{eq:final1}) determines the moments of the GFPT: 
\begin{align}\label{eq:chap50}
f^*_{1,n}(\lambda) = 1  \nonumber \\
+ \,{\frac { \left(  \left( d+1 \right) ^{n} \left(  \left( d+3 \right) ^{n+1}-1 \right) + \left( d+2 \right)  \left( d+3 \right) ^{n} \right) {d}^{2}}{\left(\left(d+1\right)^{n+1}+d-1\right)\left(d+2\right)}}\lambda  \nonumber \\
+ O(\lambda^2). \nonumber \\
\end{align}
The coefficient of $\lambda$ is the mean GFPT which matches the results found in \cite{Kozak2,Kozak}. The first and 
higher moments of the GFPT for alternative nodes can be calculated using our method. For example, the GFPT  for a trap at the mid-base of the 2-$d$ Sierpinski gasket is  
\begin{displaymath}
f_{k,n}(s) = \frac{1}{2N}\!\left(  \frac{\left( 2\,\rho_{{n}}-1 \right)}{\rho_{n}} \prod _{j=0}^{n}{\frac {2
\,\rho_{{j}}+1}{2\,\rho_{{j}}-1}}-2+{\frac {4\,\rho_{
{n-1}}-1}{\rho_{{n}}}} \right)
\end{displaymath}
where $k = 3\left(3^{n-1}+1\right)/2 - 2^{n-1}$ for $n \geq 1$, using the labeling in Figure \ref{fig:fig1}. For a discrete walker, we find $$\langle T^*  \rangle = {\frac {2\,({15}^{n})-{3}^{n}+11\,({5}^{n-1})}{{3}^{n+1}+1}}.$$

\section{Method for other renormalizable structures}\label{sec:section5}
For completeness it is useful to determine whether $f^*_{k,n}(\lambda)$ for any finitely ramified deterministic structure can be found. In the next section we are able to obtain at least one exact analytic expression.  $f^*_{k,n}(\lambda)$ for other structures can also be calculated depending on the coordination number of the nodes in the network. For the Sierpinski gasket a simple expression for  $f^*_{k,n}(\lambda)$ can be found for more than one node. This is due partly to the structure of the Sierpinski gasket, where every node is connected to $2d$ other nodes with the exception of the corners. In that case, conservation of probability simplified the solution but such a formula might not exist for other structures. However, for all finitely ramified deterministic fractals there is always a renormalization scheme which allows us to calculate $f^*_{k,n}(\lambda)$ for the case of an end node.

In the next section, we will always calculate $f_{k,n}(s)$ for node 1.
The trick in determining \eqref{eq:chap44} is to express the concentrations at the interior nodes in terms of the concentrations at the exterior (end) nodes. This is relevant when calculating \eqref{eq:chap44} in terms of the concentrations throughout the fractal, which can then be explicitly written as a function of the renormalization of the inverse FPT equation. $f^*_{k,n}(\lambda)$ can then be found from \eqref{eq:final1}. As examples we calculate $f^*_{k,n}(\lambda)$ for four different fractals. 

\begin{figure}[h]
\begin{center}
\includegraphics[width = 0.4\textwidth]{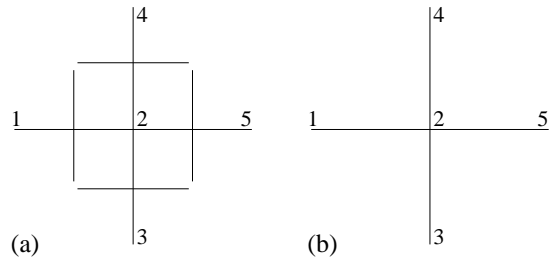}
\end{center}
\caption{Deterministic tree (a) $n = 1$ (b) $n=0$. The labeling of the nodes 1 to 5 remain the same regardless of the iteration. }
\label{fig:fig2}
\end{figure}
\subsection{Deterministic Tree}\label{sec:section55}
For the deterministic tree (Figure \ref{fig:fig2}) we calculate $f_{k,n}(s)$  for the end node 1. We can relate the concentrations at the nodes 2 to 4, for the iteration $n=0$, to the concentrations at nodes 1 and 5 in the following way:
\begin{equation}\label{eq:chap51}
 \left[ \begin {array}{c} { p_{21,0}}\\\noalign{\medskip}{ p_{31,0}}
\\\noalign{\medskip}{ p_{41,0}}\end {array} \right] 
= 
\frac{1}{2\left(2\rho_0^2 - 1\right)}
\left[ \begin {array}{cc} \rho_0 & \rho_0 \\\noalign{\medskip}1&1
\\\noalign{\medskip}1&1\end {array} \right] 
 \left[ \begin {array}{c} { p_{11,0}}\\\noalign{\medskip}{ p_{51,0}}
\end {array} \right] .
\end{equation}
Here $\rho_{n+1} = 4\rho_{n}^2 - 3$ is the renormalization of the inverse FPT equation.
We can deduce from \eqref{eq:chap51} that
\begin{equation}\label{eq:chap52}
\sum_{i=2}^{4}p_{i1,0} = \frac{\rho_0 + 2}{2(2\rho_0^2-1)}
\end{equation}
and also that
\begin{equation}\label{eq:chap53}
\sum_{i=1}^{5}\kappa_ip_{i1,0} = \frac{2\rho_0(1+\rho_0)}{2\rho_0^2-1},
\end{equation}
where $\kappa_i$ is the coordination number associated with the $i$th node. These equations are essential, as they can be used repeatedly on any iteration of the deterministic tree to express the concentrations at the interior nodes in terms of the concentrations at the nodes 1 and 5. This will now be demonstrated for iteration 1, shown in Figure \ref{fig:fig2}, which we label consistently with iteration 0 (we are not concerned with the labeling of the new nodes). We denote $\theta_{2j}$ to be the sum of the concentrations at the nodes between node 2 (i.e. the center node) and node $j$ on iteration 1. Using \eqref{eq:chap52} 
\begin{displaymath}
\theta_{21} = \frac{\rho_0 + 2}{2(2\rho_0^2-1)}(p_{11,1} + p_{21,1}).
\end{displaymath}
Repeating this process for all unlabeled nodes and summing leads to 
\begin{displaymath}
\theta_{21} + \theta_{23} + \theta_{24} + \theta_{25} = \frac{\rho_0 + 2}{2(2\rho_0^2-1)}\sum_{i=1}^{5}\kappa_ip_{i1,1}.
\end{displaymath}
The final step is recalling (\S2) that if the sum of concentrations on the LHS of \eqref{eq:chap53} gives the RHS of \eqref{eq:chap53} with $\rho_0$ for iteration 0, then one can automatically assume that the LHS of\eqref{eq:chap53} gives the RHS of \eqref{eq:chap53} with $\rho_0$ being replaced by $\rho_1$ for iteration 1. In doing so, one finds the sum of the concentrations at all non-labeled nodes on the fractal at iteration 1 is
\begin{displaymath}
\theta_{21} + \theta_{23} + \theta_{24} + \theta_{25} = \frac{\left(\rho_0 + 2\right)\left(1+\rho_1\right)\rho_1}{\left(2\rho_0^2-1\right)\left(2\rho_1^2-1\right)}.
\end{displaymath}
Furthermore, one can show that all concentrations at any node at any iteration can be expressed in terms of the concentrations at nodes 1 and 5 with the sum of concentrations at every node being
\begin{eqnarray*}
\sum_{j=1}^{N} p_{j1,n}  =  (p_{11,n}+p_{51,n}) \nonumber \\
\times  \left(1 + \sum _{m=0}^{n}\left( \frac{ \left(\rho_m 
  +2 \right)}{\left( 4\,\rho_m^{2}-2 \right)}\prod _{i=m+1}^{n}2\,{\frac{\rho_i \left(\rho_i +1\right) }
{2\,\rho_i ^{2}-1}} \right)\right). \nonumber \\
\end{eqnarray*}
We know that there are exactly $N = 4^{n+1}$ starting nodes, so $f_{1,n}(s)$ can be calculated from \eqref{eq:chap44} 
\begin{eqnarray*}
f_{1,n}(s) = \frac{1}{N} \bigg[\frac{1}{\rho_{n+1}} \nonumber \\
+ \left(1 + \frac{1}{\rho_{n+1}}\right)\sum _{m=0}^{n} \left( \frac{ \left(\rho_m +2 \right)}{\left( 4\,\rho_m^{2}-2 \right)}\prod _{i=m+1}^{n}2\,{\frac{\rho_i \left(\rho_i +1\right) } {2\,\rho_i ^{2}-1}} \right)  \bigg]. \nonumber \\
\end{eqnarray*}
Substituting for $\lambda$ \eqref{eq:final1} and taking Taylor series expansions gives
\begin{displaymath}
f^*_{1,n}(\lambda) = 1 + \left( 3({2}^{n-1})+\frac{6}{7}\,{8}^{1+n}-{\frac {17}{28}} \right) \lambda + O(\lambda^2).
\end{displaymath}

More generally, one can construct a formula for any finitely ramified determinstic fractal, provided that it has two end nodes each having a coordination number of 1. In this case, one can find the following quantities
\begin{equation}\label{eq:chap54}
y_1 \left( \rho_{{0}} \right)  =\frac{1}{p_{11,0} + p_{\nu1,0}}\sum_{\alpha} p_{\alpha1,0},   
\end{equation}
where $\alpha$ is an interior node, and 
\begin{equation}\label{eq:chap55}
y_2 \left( \rho_{{0}} \right)  =\frac{1}{p_{11,0} + p_{\nu1,0}}\sum_{j=1}^{N} \kappa_{j}p_{j1,0}. 
\end{equation}
For consistency we always let $1$ and $\nu$ represent the two end nodes of the $n$th iteration of the specific fractal. One can then show that the GFPT density for a random walker to arrive at node 1 can be expressed as 
\begin{eqnarray*}
f_{1,n}(s) = \frac{1}{N}\bigg[\frac{p_{\nu1,n}}{p_{11,n}} \nonumber \\
+ \left(1+\frac{p_{\nu1,n}}{p_{11,n}} \right)\sum _{m=0}^{n} \left( y_1 \left( \rho_{{m}} \right)  \prod _{i=m+1}^{n} y_2 \left( \rho_{{i}} \right) \right) \bigg], \nonumber \\
\end{eqnarray*}
which becomes, using the fact that $p_{\nu1,n}/p_{11,n} = \rho_{n+1}^{-1}$,
\begin{eqnarray}\label{eq:chap56}
f_{1,n}(s) = \frac{1}{N}\bigg[\rho_{n+1}^{-1} \nonumber \\
+ \left(1+\rho_{n+1}^{-1} \right) \sum _{m=0}^{n} \left( y_1 \left( \rho_{{m}} \right)  
\prod _{i=m+1}^{n} y_2 \left( \rho_{{i}}  \right)  \right) \bigg]. \nonumber \\
\end{eqnarray}
Here $\rho_{n+1}$ is the renormalization of the inverse FPT equation of the $n$th iteration of a finitely ramified deterministic fractal with two end nodes $1$ and $\nu$. To calculate $\langle T^* \rangle$ for discrete walkers we use $\rho^*_n(\lambda)|_{\lambda=0} = 1$, and the first derivative of $\rho^*_n(\lambda)$ given in table I. The method will be demonstrated on three simple examples.

\begin{table}\label{tab:table1} 
\caption{The first and second derivates of $\rho^*_n(\lambda)$, evaluated at $\lambda = 0$, for the $d$-dimensional Sierpinski gasket, the Mandelbrot-Given curve (M-G), the T-tree, the deterministic tree (D) and the hierarchical percolation model (HP).}
\begin{tabular}{|l|c|c|}
\hline
Lattice & $(\rho^*_{n})'$ & $(\rho^*_{n})''$\\
\hline
Sierpinski  & $-(d+3)^n$ & $ \left( 4\,d \left( d+3 \right) ^{n} +d+6+{d}^{2} \right)/$ \\
            &            & $(\left( d+3 \right) ^{1-n}\left(2+d\right))$ \\
\hline
M-G curve & $-22^n$ & ${22}^{n-1} \left( 359+103 \left( {22}^{n} \right) \right)/21$   \\
\hline
T-Tree  & $-6^n$ & ${6}^{n} \left(4+{6}^{n} \right)/5$  \\
\hline
D-tree & $-8^n$ & ${8}^{n} \left( 6+ {8}^{n} \right)/7 $  \\
\hline
HP model & $-10^n $ & $ 10^{n-1}\left( 7+3 \left({10}^{n}\right) \right) $  \\
\hline
\end{tabular}
\end{table}

\begin{figure}[h]
\begin{center}
\includegraphics[width = 0.35\textwidth]{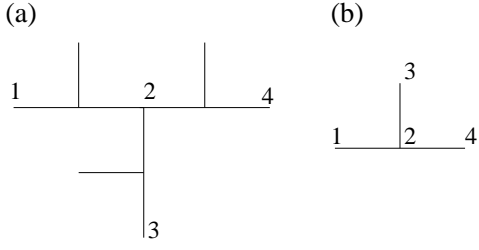}
\end{center}
\caption{The T-tree (a) $n = 1$ (b) $n=0$. The labeling of the nodes remains the same for all iterations.}
\label{fig:fig3}
\end{figure}
\subsection{T-tree}
Consider the T-tree shown in Figure \ref{fig:fig3}. The decimation procedure assumes that flux leaves only through the end nodes 1 and 4 for every iteration $n$. The concentrations at nodes 2 and 3  for the iteration $n = 0$ can be expressed in terms of the concentrations at nodes 1 and 4 in the following way:
\begin{displaymath}
\left[ \begin {array}{c} { p_{21,0}}\\\noalign{\medskip}{ p_{31,0}}
\end {array} \right] = \frac{1}{3
\,{\rho_{{0}}}^{2}-1}\left[ \begin {array}{cc} {\rho_{{0}}}&{\rho_{{0}}}
\\\noalign{\medskip} 1 & 1 \end {array} \right] 
 \left[ \begin {array}{c}  p_{11,0} \\\noalign{\medskip} p_{41,0}
\end {array} \right] .
\end{displaymath}
Hence, the quantities defined in \eqref{eq:chap54} and  \eqref{eq:chap55} are
\begin{displaymath}
y_2\left(\rho_0\right) = 3\rho_{0}y_1\left(\rho_0\right) = 3\,{\frac {\rho_{{0}} \left( 1+
\rho_{{0}} \right) }{3\,{\rho_{{0}}}^{2}-1}}.
\end{displaymath}
We use the renormalization of the inverse FPT equation $\rho_{n+1} = 3\rho_{n}^2 - 2$ and $N=3^{n+1}$ in equation \eqref{eq:chap56}, to find 
\begin{align}\label{eq:chap77}
f_{1,n}(s) = \frac{1}{N} \bigg[{\rho_{{n+1}}}^{-1}  \nonumber \\ 
+ \left( 1+{\rho_{{n+1}}}^{-1} \right)\sum _{m=0}^{n} \left(  \frac{\left( 1+\rho_{{m}} \right)}{3\,{\rho_{{m}}}^{2}-1 } \prod _{i=m+1}^{n}3\,{\frac { \left( 1+\rho_{{i}} \right) \rho_{{i}}}{3\,{\rho_{{i}}}^{2}-1}}  \right) \bigg]. \nonumber\\
\end{align}
Substituting for $\lambda$ \eqref{eq:final1} in \eqref{eq:chap77} and taking Taylor series expansions gives
\begin{displaymath}
f^*_{1,n}(\lambda) = 1 + \left( \frac{{2}^{n+2}}{3}+{\frac {4}{5}}\,{6}^{n+1}-\frac{7}{15} \right) \lambda + O(\lambda^2).
\end{displaymath}
The mean coincides with the $\langle T^* \rangle$ for a trap at the center (node 2) of the T-tree of iteration $n-1$ (replace $n$ by $n-1$ in the above formula), which agrees with \cite{Agliara}. 

For the next two examples we just state $N$, the renormalization of the inverse FPT equation, and the two essential quantities $y_1(\rho_0),y_2(\rho_0)$.

\begin{figure}[h]
\begin{center}
\includegraphics[width = 0.48\textwidth]{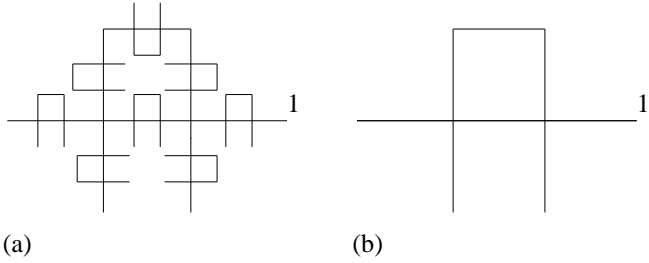}
\end{center}
\caption{Mandelbrot-Given curve (a) $n=1$  (b) $n=0$.}
\label{fig:fig4}
\end{figure}

\subsection{Mandelbrot-Given curve}
The Mandelbrot-Given curve is shown in (Figure \ref{fig:fig4}). We find that
$N = \left(6\left(8^{n+1}\right) + 1\right)/7$, $\rho_{n+1} = -21\,\rho_n+13/2\,{\rho_n}^{-1}-1/2\,{\rho_n}^{-3}+16\,{\rho_n}^{3}$ ,
\begin{displaymath}   
y_1(\rho_0) = \frac{2\, \rho_0^{2}+2\,\rho_0 -1 }{ 8\,\rho_0^{3}-6\,\rho_0^{2}-2\,\rho_0 +1}
\end{displaymath}
and
\begin{displaymath}
y_2(\rho_0) =   \frac{2\,\rho_0  \left( 4\,\rho_0^{2}+\rho_0 -1 \right)}{8\,\rho_0 ^{3}-6\, \rho_0^{2}-2\,\rho_0 +1}.
\end{displaymath}
Substituting these expressions into equation \eqref{eq:chap56}, we find in the usual way that
\begin{eqnarray*}
f^*_{1,n}(\lambda) = 1 + 4/7\frac{1518(176^n)-39(8^n)+187(22^n)}{1+48(8^{n})}\lambda \nonumber \\
+ O(\lambda^2). \\
\end{eqnarray*}
The coefficient of $\lambda$ is the mean GFPT for a random walker to reach node 1 of the $n$th iteration of the Mandelbrot-Given curve.

\begin{figure}[h]
\begin{center}
\includegraphics[width = 0.35\textwidth]{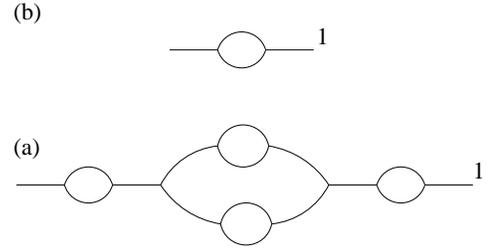}
\end{center}
\caption{Hierarchicial percolation model (a) $n = 1$ (b) $n=0$.}
\label{fig:fig5}
\end{figure}
\subsection{Hierarchical percolation model}

The hierarchical percolation model shown in Figure \ref{fig:fig5} has the properties $N = (2({4}^{n+1})+1)/3$, $\rho_{n+1} = 
 \rho_{n}(9\rho_{n}^2 - 7)/2$ and 
\begin{displaymath}
y_2(\rho_0) = (3\rho_0 + 1)y_1(\rho_0) = \frac{(3\rho_0 + 1)}{3\rho_0-2}.
\end{displaymath}
Using these results in equation \eqref{eq:chap56}, transforming to $\lambda$, and taking Taylor series expansions gives 
\begin{displaymath}
f^*_{1,n}(\lambda) = 1 + 2\,{\frac {{10}^{n+1}+30({40}^{n})-{4}^{n}}{1+2({4}^{n+1})}}\lambda + O(\lambda^2).
\end{displaymath}
Here, as in all other examples, the coefficient of $\lambda$ is the mean of the GFPT for node 1 of the hierarchical percolation model. 

\section{Scaling of higher order moments}
To find higher moments of the GFPT, it is necessary to calculate higher derivatives of $\rho^*_n(\lambda)$. We have given the relevant results in Table I. As an example, we calculate the second moment of the GFPT on a T-Tree. 
From the coefficient of $\lambda^2$ in the Taylor expansion of $f^*_{n}(\lambda)$ (\ref{eq:chap77} in \ref{eq:final1}) we find that the second moment is given by
\begin{equation}\label{eq:chap57}
\langle \left(T^*\right)^2 \rangle = {\frac {72}{5}}\,{12}^{n}-{\frac {16}{15}}\,{2}^{n}+{\frac {8928}{175}
}\,{36}^{n}-{\frac {208}{25}}\,{6}^{n}+{\frac {53}{175}}.
\end{equation}
It is interesting to examine how the first and second moments, expressed as a function of the number of nodes $N_T = N+1$, scale with the spectral dimension $\bar{d}$ \cite{Orbach}. Usually, $\bar{d}$ is determined through the analysis of the Green's function at the origin $G_0(t)$, which is known to scale as
\begin{displaymath}
G_0(t) \sim t^{-\bar{d}/2}.
\end{displaymath}
In \cite{Kozak,Kozak2}, the mean GFPT expressed in terms of the total number of nodes $N_T$ on the Sierpinski gasket was related  to the spectral dimension $\bar{d} = 2\ln(d+1)/\ln(d+3)$. This was done by re-arranging the formula for the number of nodes on a $d$-dimensional gasket, $N_T = N+1 =\left(d+1\right)\left(\left(d+1\right)^n + 1\right)/2$, to set $(d+1)^n = N_T\frac{2}{d+1}-1$ and $(d+3)^n = (N_T\frac{2}{d+1}-1)^{2/\bar{d}}$. Substituting these values into \eqref{eq:chap50} and taking the limit for large $N_T$ gives $\langle T^* \rangle \sim N_T^{2/\bar{d}}$. 

We perform a similar analysis for the second moment of the GFPT given in \eqref{eq:chap57}. There are $N_T = 3^{n+1}+1$ nodes available on the $n$th iteration of the T-fractal. By replacing $3^{n}$ by   $\left( N_T-1 \right)/3$ and $2^{n}$ by $\left(\left(N_T-1\right)/3\right)^{2/\bar{d}-1}$ in the expression for the second moment of the GFPT \eqref{eq:chap57} we find
\begin{align}
\langle \left(T^*\right)^2 \rangle  = 6/5\, \left( N_T-1 \right) ^{4/{\bar{d}}-1}-{\frac {8}{15}}\, \left( N_T-1
 \right) ^{2/\bar{d}-1}+ \nonumber\\
{\frac {248}{175}}\, \left( N_T-1 \right) ^{4/\bar{d}}-{\frac {104}{75}}\, \left( N_T-1 \right) ^{2/\bar{d}}+{
\frac {53}{175}}. \nonumber\\
\end{align}
Asymptotically we find that $ \langle \left(T^*\right)^2 \rangle \quad \sim \quad N_T ^{4/\bar{d}}$. Since we have an exact expression for the mean and second moment we can show that the variance scales as
\begin{equation}\label{eq:chap61}
\textrm{Var}(T^*) \quad \sim \quad N_T^{4/\bar{d}}.
\end{equation}
All other structures considered above yield the same asymptotic behavior.

\section{Conclusion}
We have used the properties of the renormalization of the first passage time equation to find the first passage time moment generating function for a variety of fractals. From this, we were able to derive simple expressions for the first and second moments of the global first passage time.  

\begin{figure}[h]
\begin{center}
\includegraphics[width = 0.5\textwidth]{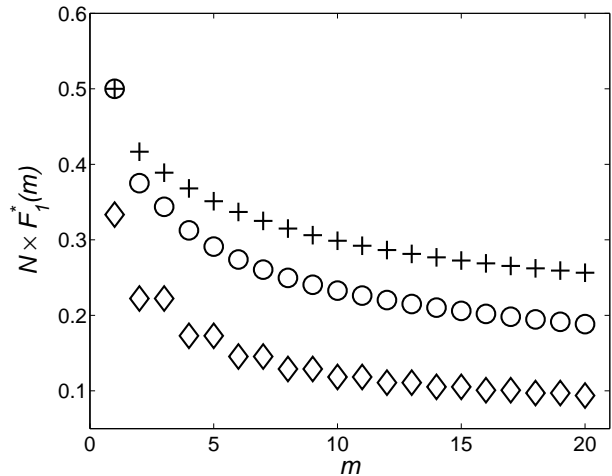}
\end{center}
\caption{The normalized probability for a random walker to arrive at a trap at node 1 on the $m$th jump for the $10$th iteration of the Sierpinski gasket ($d=2$) ($\circ$), Sierpinski gasket ($d=3$) (+)  and the hierarchical percolation model ($\diamond$). The values are calculated from the Taylor series of the GFPT generating function $f^*_{1}(\ln(z))$. A symbolic algebra package was used.}
\label{fig:fig6}
\end{figure}
Our results can be applied to a diffusion controlled reaction experiment of the form $A + B \rightarrow B$. Here $A$ is a particle which performs a random walk on the structure until it is absorbed by a single immobile particle $B$. If there are an infinite number of $A$ particles then the reaction rate at node $k$ is the deterministic function $F_k^*(m)$. 
This reaction rate, at node $k=1$, for the gasket in two and three dimensions, and the hierarchical percolation model, is shown for small time in Figure \ref{fig:fig6}. The first and second moment we have calculated are the moments of the reaction rate. An alternate experiment would involve a single walker in which case the expected time of completion is $\langle T \rangle$ with the outcome being extremly variable, the variance being given by $\textrm{Var}(T^*) \sim N_T^{4/\bar{d}}$.

Since the variance of the first passage time scales as \eqref{eq:chap61}, we relate the variance to the mean by calculating the reduced moment $R(N_T) = \textrm{Var}(T^*)^{1/2}/\langle T^* \rangle$. This type of approach has been used in the study of the variance of the range of an $n$-step random walk \cite{Benoit,Kopelman}. As in \cite{Benoit} we expect that $\lim_{N_T \rightarrow \infty} R(N_T)$ exists and remains finite, for every fractal structure with a spectral dimension of $\bar{d}<2$. Using \eqref{eq:chap49}, we find the reduced second moment for the Sierpinski gasket to be
\begin{displaymath}
\lim_{N_T \rightarrow \infty} R(N_T) = \left( 1+{\frac {8}{ \left( d+3 \right)  \left( d+4 \right) d}} \right )^{1/2},
\end{displaymath} 
which decreases monotonically with $d$.

\begin{acknowledgments}
The manuscript has been considerably improved by numerous detailed comments made by the referees. 
\end{acknowledgments}

\appendix
\section{Corner concentrations}\label{sec:append1}
Here we show how the concentration at the apex and the corners of the $n$th iteration of the $d$-dimensional Sierpinski gasket can be calculated if a walker is released at the apex. We define $p_{c(i)1,n}$ to be the concentration at the $i = 1..d$  corners of the gasket. Recall that the concentration at the apex is given by $p_{11,n}$. The flux-concentration matrix has the simple (circulant) form 
\begin{displaymath}
\left(\begin{array}{cc} 
1 \\
0\\
\vdots \\
0
\end{array}\right)
=
\left( \begin{array}{cccc}
da_n & -b_n & \ldots & -b_n \\
-b_n & da_n & -b_n & \vdots \\
\vdots & \vdots& \ddots & \vdots \\
-b_n  & \ldots & -b_n & da_n \\
\end{array} \right)
\left(\begin{array}{cc} 
p_{11,n} \\
p_{c(1)1,n} \\
\vdots \\
p_{c(d)1,n}\\
\end{array}\right).
\end{displaymath}
This is the natural extension of equation \eqref{eq:chap9}, with $a_n$ and $b_n$ defined in equation \eqref{eq:chap28}. The solution of the equations are 
\begin{equation}\label{eq:append2}
p_{c(i)1,n} = \frac{b_n}{d\left(a_n - b_n\right)\left(da_n + b_n\right)} = \frac{1}{db_n\left(\rho_n - 1\right)\left(d\rho_n + 1\right)}    
\end{equation}
for $i = 1,\ldots,d$, and  
\begin{equation}\label{eq:append3}
p_{11,n} = \frac{\left(da_n - (d-1)b_n\right)}{d\left(a_n - b_n\right)\left(da_n + b_n\right)} = \frac{\left(d\rho_n - (d-1)\right)}{db_n\left(\rho_n - 1\right)\left(d\rho_n + 1\right)}.
\end{equation}
If we use the renormalization of first passage time equation $\rho_{n}\left(d\right) = 2d\rho^2_{n-1} -3\left(d-1\right)\rho_{n-1} + \left(d-2\right)$ (See \eqref{eq:chap29}) on the term $\left(\rho_n - 1\right)$ in $p_{11,n}$ \eqref{eq:append3}, we find that 
\begin{displaymath}
p_{11,n} = \frac{\left(d\rho_n - (d-1)\right)}{db_n\left(2d\rho_{n-1} - d+3\right)\left(\rho_{n-1} -1\right)\left(d\rho_n + 1\right)}.
\end{displaymath}
Using $$b_{n} = {\frac { \left(  d\rho_{n-1}+  1\right) b_{n-1} }{ \left( d\,\rho_{n-1}- \left( d-1 \right)\right)  \left( 2d\, \rho_{n-1}+\left(3-d\right)\right)}},$$ we obtain
\begin{displaymath}
p_{11,n} = \frac{\left( d\,\rho_{n-1}- \left( d-1 \right)\right)}{db_{n-1}\left(\rho_{n-1} -1\right)\left(d\rho_{n-1} + 1\right)}\frac{\left(d\rho_{n} - (d-1)\right)}{\left(  d\rho_{n}+  1\right)}.
\end{displaymath}
Note the similarities with equation \eqref{eq:append3}. If we repeat the above procedure we get the result
\begin{equation}\label{eq:append4}
p_{11,n} = \frac{\sinh \left(\sqrt {s} \right) }{d\sqrt {s} \left( \cosh \left( \sqrt {s} \right) -1 \right)}\prod _{j=0}^{n}\frac{d\rho_j -(d-1)}{d\rho_j +1}.
\end{equation}
The concentration at the corner points, defined in equation \eqref{eq:append2}, can be found using the same procedure used above. The result can be expressed as $p_{c(i)1,n} = p_{11,n}/(d\rho_n -(d-1)).$
From the two preceeding results it can be shown that $p_{11,n} + dp_{c(i)1,n} = p_{11,n-1}$ which agrees with the more general result given in \S\ref{sec:section3}.

\section{Continuum to discrete transformations}\label{sec:append2}
The results we derive for a continuous space/continuous time random walker can be applied to a discrete walker by simple transformations. For a discrete walker the probability flux and probabilities at ends 1 and 2 of a bar are related by 
\begin{eqnarray*}
F_1^*(m) = \frac{P_1^*(m)}{\kappa_1}  - \frac{P_2^*(m-1)}{\kappa_2}  \\ \nonumber
F_2^*(m) = \frac{P_2^*(m)}{\kappa_2} - \frac{P_1^*(m-1)}{\kappa_1}   \\ \nonumber
\end{eqnarray*}
where  $m\geq 0$ and $\kappa_i$ is the coordination number of the $i$th node. These equations are quite formal and are easier to understand when they are combined for a network.

Multiplying both equations by $z^n$ and summing from $n=0,\ldots,\infty$ leads to 
\begin{equation}\label{eq:append31} 
\left[ \begin {array}{c} {\it \hat{f}_1}\\\noalign{\medskip}{\it \hat{f}_2}
\end {array} \right] = \left[ \begin {array}{cc} 1/\kappa_1 &  - z/\kappa_2 
\\\noalign{\medskip} -z/\kappa_1 & 1/\kappa_2 \end {array} \right]  
\left[ \begin {array}{c} {\it \hat{p}_1}\\\noalign{\medskip}{\it \hat{p}_2}\end {array} \right], 
\end{equation}
where $\hat{p}_i(z) = \sum_{m=0}^{\infty}P^*(m)z^{m}$ is the probability generating function of $P^*(m)$.  
We have used the fact that $\sum_{i=0}^{\infty}P^*_i(m-1)z^{m} = P^*_i(-1) + z\sum_{i=0}^{\infty}P^*_i(m)z^{m}$ and define $P^*_i(-1) = 0$. Equation \eqref{eq:append31} is the direct analog of \eqref{eq:chap4}.

The transformations we need to derive are most easily shown by example. Using \eqref{eq:append31} for the first iteration of the 2-$d$  Sierpinski gasket gives;
\begin{equation}\label{eq:append32}
\left[ \begin {array}{c} { \hat{f}_1}\\\noalign{\medskip}{ \hat{f}_2}
\\\noalign{\medskip}{ \hat{f}_3}\\\noalign{\medskip}{ \hat{f}_4}
\\\noalign{\medskip}{ \hat{f}_5}\\\noalign{\medskip}{ \hat{f}_6}
\end {array} \right] 
= 
\left[ \begin {array}{cccccc} a_0&-\frac{b_0}{4}&-\frac{b_0}{4}&0&0&0\\\noalign{\medskip} -\frac{b_0}{2} & a_0 & -\frac{b_0}{4} & - \frac{b_0}{2} & -\frac{b_0}{4} & 0\\\noalign{\medskip} - \frac{b_0}{2} &- \frac{b_0}{4} & a_0 &0 &-\frac{b_0}{4}&-\frac{b_0}{2} \\\noalign{\medskip} 0 &-\frac{b_0}{4} &0
&a_0&-\frac{b_0}{4}&0\\\noalign{\medskip}0&-\frac{b_0}{4}&-\frac{b_0}{4} & -\frac{b_0}{2} & a_0&-\frac{b_0}{2} \\\noalign{\medskip}0&0&-\frac{b_0}{4}
&0&-\frac{b_0}{4}& a_0\end {array} \right] 
\left[ \begin {array}{c} { \hat{p}_1}\\\noalign{\medskip}{ \hat{p}_2}
\\\noalign{\medskip}{ \hat{p}_3}\\\noalign{\medskip}{ \hat{p}_4}
\\\noalign{\medskip}{ \hat{p}_5}\\\noalign{\medskip}{ \hat{p}_6}
\end {array} \right], 
\end{equation}
where $a_0 = 1$ and $b_0 = -z$. We release a walker at site 1  at $t = 0$ and place a trap at site 6 with probability conservation at other nodes. This means ${\bf \hat{p}}^T= (\hat{p}_1,\hat{p}_2,\hat{p}_3,\hat{p}_4,\hat{p}_5,0)$ and ${\bf \hat{f}}^T=(1,0,0,0,0,\hat{f}_6)$, where $-\hat{f_6}$ is the probability generating function of the first passage time. The matrix equation given above can also be directly derived from the equations governing the probability of a discrete walker on the gasket (i.e. $P^*_1(m+1) = P^*_2(m)/4 + P^*_3(m)/4$ for node 1, etc). The initial and boundary conditions are represented in our flux vector. 

If we write the matrix equation \eqref{eq:append32} as ${\bf \hat{f}} = A {\bf \hat{p}}$ and multiply by $g = \textrm{tanh}(\sqrt{s})/\sqrt{s}$, we obtain 
\begin{equation}\label{eq:append33}
g{\bf \hat{f}} = gA {\bf \hat{p}}.
\end{equation}
Rescaling ${\bf \hat{p}}$ as  
$$
{\bf \hat{p}}^{T} = (\frac{2\hat{p}'_1}{g},\frac{4\hat{p}'_2}{g},\frac{4\hat{p}'_3}{g},\frac{2\hat{p}'_4}{g},\frac{4\hat{p}'_5}{g},0) 
$$
transforms the matrix equation \eqref{eq:append33} to $g{\bf \hat{f}} = A' {\bf \hat{p}'}$. Because of our rescaling of  ${\bf \hat{p}}$, the matrix $A'$ is now identical to the flux-concentration matrix defined in \eqref{eq:chap12} if $z$ is replaced by $\textrm{sech}(\sqrt{s})$. Comparing the variables in both problems shows how first passage time and probability generating functions for discrete walkers are related to their continuum counterparts.

More generally, for the quantities considered in this paper, the above example establishes the following transformations;
\begin{displaymath}
p_{ij}(s) = \frac{1}{\kappa_i}\frac{\tanh(\sqrt{s})}{\sqrt{s}}\hat{p}_{ij}(z)\Big|_{z = \textrm{sech}(\sqrt{s})},
\end{displaymath} 
and $f_{ij}(s) = \hat{f}_{ij}(\textrm{sech}(\sqrt{s})).$ Recall $\hat{f}$ is the probability generating function of the first passage time. In order to relate $\hat{f}$ to the moment generating function $f^*$, we make the further substitution $ f^{*}(\lambda) = \hat{f}(e^{\lambda})$. Thus the quantities in the main text are given by 
\begin{displaymath}
f^{*}_{kj}(\lambda) = \hat{f}_{kj}(e^{\lambda}) = f_{kj}(\textrm{arcsech}(e^{\lambda})^2)
\end{displaymath}
which is the transformation used throughout the paper. 

We have not seen these transformations before, but they are analogous to the Montroll-Weiss transformations \cite[Page 254, Eq. 5.51]{Hughes}.  which relate the properties of a continuous time discrete space random walker to those of a discrete time discrete space random walker.

\end{document}